\documentclass[sigplan,nonacm,screen]{acmart}
\usepackage[T1]{fontenc}

\title{Modeling Utilization to Identify Shared-memory Atomic Bottlenecks}
\author{Rongcui Dong}
\affiliation{%
\institution{University of Rochester}
\city{Rochester}
\state{New York}
\country{USA}}

\author{Sreepathi Pai}
\affiliation{%
\institution{University of Rochester}
\city{Rochester}
\state{New York}
\country{USA}}

\usepackage{natbib}
\usepackage[ruled,vlined]{algorithm2e}
\usepackage{subcaption}
\usepackage{listings}
\lstdefinestyle{mystyle}{
    basicstyle=\ttfamily\small,
    numbers=left,
    numbersep=2pt,
}
\usepackage{booktabs}
\usepackage[inline]{enumitem}
\usepackage{mathtools}
\usepackage{microtype}

\acmConference{GPGPU'2025}{March 1}{Las Vegas, NV}
\begin{document}
\begin{abstract}
    Performance analysis is critical for GPU programs with data-dependent behavior, but models like Roofline are not very useful for them and interpreting raw performance counters is tedious.
    In this work, we present an analytical model for shared memory atomics (\emph{fetch-and-op} and \emph{compare-and-swap} instructions on NVIDIA Volta and Ampere GPU) that allows users to immediately determine if shared memory atomic operations are a bottleneck for a program's execution.
    Our model is based on modeling the architecture as a single-server queuing model whose inputs are performance counters.
    It captures load-dependent behavior such as pipelining, parallelism, and different access patterns.
    We embody this model in a tool that uses CUDA hardware counters as parameters to predict the utilization of the shared-memory atomic unit.
    To the best of our knowledge, no existing profiling tool or model provides this capability for shared-memory atomic operations.
    We used the model to compare two histogram kernels that use shared-memory atomics.
    Although nearly identical, their performance can be different by up to 30\%.
    Our tool correctly identifies a bottleneck shift from shared-memory atomic unit as the cause of this discrepancy.
\end{abstract}

\maketitle

\section{Introduction}\label{sec:intro}

The GPU is known for its ability to execute massively data-parallel programs efficiently~\cite{owens2008gpu,navarro2014survey}.
This makes it indispensable for problems such as machine learning~\cite{awan2017depth,li2017survey}, physical simulation~\cite{hermann2010multi,liang2018gpu,jespersen2010acceleration}, bioinformatics~\cite{vouzis2011gpu}, and many others.
To achieve high performance, programs must effectively use the GPU's memory hierarchy.
In particular, making good use of the on-chip scratchpad memory known as \textit{shared memory} on NVIDIA GPUs is key to avoid the high latency cost imposed by other parts of the GPU memory hierarchy.

A common use of shared memory is to optimize algorithms in tight loops~\cite{egielskiMassiveAtomicsMassive2014,GPUProTipHistogram2015,harrisOptimizingParallelReduction2007,CUDAProTipFilter2014,nylandUnderstandingUsingAtomica},
but using shared memory effectively for programs with irregular parallelism or those that have data-dependent access patterns is hard.
Of particular interest to us in this work are programs with irregular and data-dependent parallelism that use shared memory atomics to reduce the cost of global memory atomics, such as histogram~\cite{konstantinidisPracticalPerformanceModel2015,egielskiMassiveAtomicsMassive2014,gomez-lunaPerformanceModelingAtomic2013}, sorting~\cite{cudacccl}, sparse matrix multiplication~\cite{wang2010optimizing}, non-deterministic finite automata~\cite{liuWhyGPUsAre2020}, etc.
In these programs, a large problem is divided so that each sub-problem fits inside shared memory to compute partial results, which are then combined in global memory.
Using shared-memory atomics leads to lower contention and higher performance.

Since shared-memory atomic operations can also become the bottleneck of a GPU program, it is desirable to understand how much time is spent on these operations.
Unfortunately, no existing GPU profiler or performance model currently measures this time.
The CUDA toolkit provides call graph timing, performance counters, memory analysis, and Roofline charts~\cite{cuda_profiler_guide}.
Performance counters for shared-memory atomic operations include count, peak throughput, and peak parallelism.
However, these values do not easily translate to the actual cost, because the time required for shared-memory atomic operations is not publicly documented and is also load dependent.
For instance, as we will show in experiments, the average time required for atomic operations can vary more than ten times depending on their launch and access patterns.
The Roofline charts only analyze bottlenecks for arithmetic and memory operations and do not analyze atomic operations~\cite{cuda_profiler_guide,NsightKernelProfilingGuide}.
Therefore, no current tools or models reveal the utilization of shared-memory atomic units for programs that are data-dependent with irregular access patterns.

In this work, we contribute:
\begin{enumerate*}
    \item an operational single-server queue-based performance model;
    \item a tool that uses the performance model to estimate utilization of shared-memory atomic units in real programs using quantities derived from hardware performance counters.
    \item a case study of two histogram variants on two different GPUs that reveals bottlenecks which cause up to 30\% performance difference. %
\end{enumerate*}

This paper is organized as follows. %
In Section~\ref{sec:characterize}, we describe the atomic instructions of the NVIDIA Titan V GPU and the Ampere A6000 GPU that we model in this work. %
In Section~\ref{sec:ssqm}, we give a brief introduction to the operational method of queuing analysis, build a single-server queuing model for these instructions, and find hardware performance counters for obtaining basic quantities used as parameters to the model.
In Section~\ref{sec:histogram}, we perform a case study on two histogram kernels using our performance model.
In Section~\ref{sec:related}, we describe some related works and compare them with our work.
We %
conclude in Section~\ref{sec:conclusion}.

\section{Shared Memory Atomic Instructions}\label{sec:characterize}

Atomic operations allow data to be read, modified, and written in a single step.
These operations are critical in multithreaded programming models, because they prevent intermediate states from being observed by other threads.
An example of fetch-and-op (FAO) would be fetch-and-add, which performs the memory read, adds a constant, and store the sum in one operation.
Other operations are subtraction, exchange, minimum, maximum, increment, decrement, bitwise AND, bitwise OR, and bitwise XOR.
The compare-and-swap (CAS) operation conditionally replaces a value if it matches a specified value, returning the old value.
FAO is a cheaper arithmetic operation and is used whenever possible.
CAS instructions may be generated when FAO is insufficient, such as floating point atomic operations.

In the Ampere architecture, NVIDIA introduced a new atomic instruction, \texttt{ATOMS.POPC.INC}, to conditionally replace \texttt{ATOMS.ADD} instructions.
\texttt{ATOMS.POPC.INC} remain undocumented, but experiments show that it increments a shared memory variable by the number of active threads in a warp (i.e. \emph{population count increment}, or \texttt{POPC.INC}).
This instruction is generated when an atomic increment does not use its return value, and it is cheaper to execute than FAO instructions.
As we show later in Section~\ref{sec:histogram}, histogram kernels generate this instruction.

\section{Single-Server Queuing Model}\label{sec:ssqm}

Queuing models are used to analyze the performance of computer systems~\cite{denning_operational_1978,kleinrock1975theory} by modeling it as a network of \emph{servers} servicing \emph{jobs}.
A job is a unit of work.
Each server is an abstract entity with an input queue and services jobs sequentially with some service time $S$.
If there are multiple servers in the network, a job requests service from only one server at a time.

We analyze our queuing network
using the operational approach~\cite{denning_operational_1978} %
because we found that most (but not all) of the quantities required can be measured directly using GPU performance counters.%

Importantly, our model does not have to exactly represent the internal architecture of the shared memory atomic unit.
Even though we know through microbenchmarking that it is pipelined and can process multiple instructions,
we found it sufficient to model the performance of the shared-memory atomic unit using a load-dependent, single-server queuing model.
Thus, there is only one server whose service time for a job varies with job class and the number of jobs being serviced and in queue.

\subsection{Model Parameters}

In our single-server model, a \emph{job} is defined as one warp-instruction.
For our model, we need the \emph{service time} $S$, the average time (in cycles) required for a job to complete, not including queuing delay.
Parallelism and pipelining in the hardware implementation make the service time dependent on \textit{load}, denoted $n$, which is the number of jobs in queue or receiving service at the server.
Due to serialization caused by bank conflicts, the number of active threads, $e$, also affects the time required for an atomic operation: a warp-instruction with 16 threads accessing the same bank takes less time than one with 32 active threads.
To capture this effect, service time also depends on $e$.
Because FAO and CAS instructions share the same pipeline and have different latencies, service time also depends on the number of CAS operations, $c$.

To illustrate these quantities: if there are 3 other atomic warp-instructions in queue at the time when a \verb|atomicAdd| warp-instruction is being serviced, $n$ is $4$.
Additionally, if the warp-instruction being serviced involves 16 threads accessing the same location, its $e$ is $16$.
If one of the warp-instructions is a CAS instruction, $n$ is still $4$ but $c$ is $1$.

Thus, to use a single-server model, we need to obtain the value of $S(n, e, c)$ for all possible $n$, $e$, and $c$ to capture both the effects of load, as well as the number of active threads.

\subsection{Measuring Model Parameters}

A microbenchmark that simply times atomic instructions %
does not measure $S$, because the measurement is the \emph{response time}, which includes both \emph{service time} and \emph{queuing time}.
Therefore, we must derive $S$ from other quantities.
We use a variation of the operational law \emph{Mean Service Time Between Completions}~\cite{denning_operational_1978}, $S(n,e,c) = T(n,e,c)/C(n,e,c)$, to obtain $S(n,e,c)$.
$T$ is the \emph{total time} from first arrival to last completion.
$C$ is the \emph{completion count}, the total number of jobs completed.
In our microbenchmark, we control the number of job arrivals, $A$, and measure the total time $T(n,e,c)$.
Since we issue all atomic operations at once, all arrivals must be queued up initially, so $n = A$.
Since our measurement period includes all arrivals and completions, satisfying \emph{job flow balance}, $C(n,e,c) = A$.
Therefore, we can derive $S(n,e,c) = T(n,e,c) / n$.
Since each warp contains 32 threads, $e$ can be most 32.
In Volta GPUs, each SM can run up to 64 warps~\cite[Table~15]{cuda_c_programming_guide}, so $n$ can be at most 64.
Ampere GPUs can have $n$ up to 48.

\begin{figure}
    \centering
    \includegraphics[width=\linewidth]{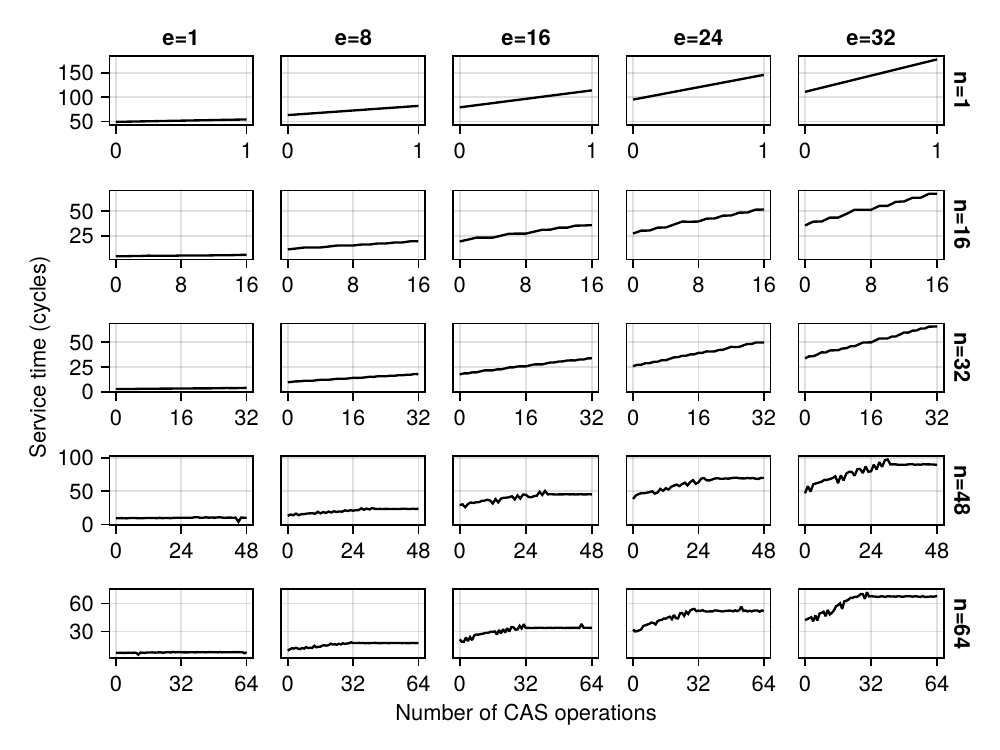}
    \Description{A 5 by 5 grid of line plots showing number of CAS vs measured total time. Subplot rows vary the number of warps from 1 to 64 and columns varying the number of active threads per warp from 1 to 32. Plots show trend of linear increase when number of warps are fewer than 32, and saturates when greater. Service time decreases with more warps, increases with more active threads, and increases with more CAS.}
    \caption{Single-server model parameters of Titan V. In each subplot, X-axis is $c$ and Y-axis is $S$. Each row corresponds to one $n$ and each column correspond to one $e$. To reduce clutter, we only show some values for $n$ and $e$ but we measure for all possible integral $n$ and $e$ values.}\label{fig:ssm}
\end{figure}

Figure~\ref{fig:ssm} shows the service time of the single-server queuing model on the Titan V GPU with Volta architecture.
Each row is a different queue length, while each column is a different number of active threads per warp.
In each subplot, X-axis is $c$ and Y-axis is $S$.
For instance, the second row ($n = 16$) has 16 warp-instructions, where $c\le n$ warps execute CAS instructions and the rest execute FAO instructions, and each warp has $e \le 32$ active threads.
Each data point shows the time interval $T$ between the start time of the first thread and the finish time of the last thread divided by 16.

Overall, service time $S$ decreases as $n$ increases due to higher parallelism, and increases as $e$ increases. %
Mixing FAO and CAS instructions (i.e. changing $c$) gives a roughly linear change in service time, except when $n > 32$ (i.e. when more than one block runs on the same SM), saturating to the CAS service time in that case.

Although we measure $T$ (and thus $S$) for integral values of $n$ and $e$, the inputs to the model can be non-integral (Table~\ref{tab:derived-quantities}).
In these cases, we interpolate $T$ using linear interpolation (Equation~\ref{eq:lerp}) and then compute $S$ from $T$ (Equation~\ref{eq:s}).

\begin{align}
    \hat{T}(n,e,c) & = \begin{cases}    T_{sampled}(n,e,c) & \text{if $0 < n \le n_{max}$} \\
                 0                  & \text{if $n = 0$}
                       \end{cases} \label{eq:t-int} \\
    T(n,e,c)       & = \text{linearInterpolation($\hat{T}$, $n$, $e$, $c$)} \label{eq:lerp}                      \\
    S(n,e,c)       & = \frac{T(n,e,c)}{n} \label{eq:s}
\end{align}

We only need to collect $T(n, e, c)$ (and thus $S(n,e,c)$) once per GPU model.
Since $T(n, e, c)$ does not reveal any hardware implementation details, we suggest that manufacturers provide such a table as part of their specifications instead of just providing the values corresponding to peak throughput as is done in current documentation. %

\subsection{Applying the Model}

\begin{table*}
    \centering
    \caption{Basic operational quantities}
    \label{tab:counters}
    \begin{tabular}{cccl}\toprule
        Quantity    & Description              & Tool   & Counter                                                   \\\midrule
        $O$         & Total atomic operations  & NCU    & \texttt{smsp\_\_l1tex\_\ldots\_mem\_shared\_op\_atom.sum} \\
        $N_f^{(i)}$ & FAO jobs on SM $i$       & NVProf & \texttt{shared\_atom}                                     \\
        $N_c^{(i)}$ & CAS jobs on SM $i$       & NVProf & \texttt{shared\_atom\_cas}                                \\
        $T^{(i)}$   & Total time on SM $i$     & NVProf & \texttt{active\_cycles}                                   \\
        $o^{(i)}$   & Avg. occupancy on SM $i$ & NVProf & \texttt{achieved\_occupancy}                              \\
        \\\bottomrule
    \end{tabular}
\end{table*}

\begin{table}
    \centering
    \caption{Derived operational quantities}
    \label{tab:derived-quantities}
    \begin{tabular}{ccc}\toprule
        Quantity        & Derivation                                 & Description                 \\\midrule
        $N^{(i)}$       & $N_f^{(i)} + N_c^{(i)}$                    & Total atomic jobs on SM $i$ \\
        $\hat{n}^{(i)}$ & $o^{(i)}\text{WarpsPerSM}$                 & Avg. parallelism on SM $i$  \\
        $e$             & $O / \sum_i N^{(i)}$                       & Avg. active threads per job \\
        $c^{(i)}$       & $n\frac{N_c^{(i)}}{N_c^{(i)} + N_f^{(i)}}$ & Avg. queued CAS on SM $i$   \\
        $B^{(i)}$       & $N \times S(\hat{n}, e, c)$                & Busy time on SM $i$         \\
        $U^{(i)}$       & $B^{(i)} / T^{(i)}$                        & Utilization on SM $i$       \\\bottomrule
    \end{tabular}
\end{table}

To use the model on real programs, we need hardware performance counters that correspond to operational quantities.
Table~\ref{tab:counters} shows the list of basic quantities, tools used for measurement, and the counters from which they are measured.
Table~\ref{tab:derived-quantities} shows how we use these counters to obtain the quantities used in our queuing model to estimate utilization.

All quantities in Table~\ref{tab:counters} collected by NVProf are collected per-SM, denoted with superscript $(i)$ for SM $i$, while NCU reports totals across all SMs.
NCU or NVProf are CUDA profiling tools. %
NVProf uses the old CUPTI Metrics API, while NCU uses the new CUPTI Profiling API~\cite{cuda_cupti}.
The Metrics API supports fewer performance counters, but can collect per-SM data.
The Profiling API supports more performance counters, but can only aggregate quantities across all SMs, providing \emph{minimum}, \emph{maximum}, \emph{sum}, and \emph{average}.
Because the single-server model captures behavior of a single SM, we try to use per-SM quantities collected from NVProf whenever possible.
The total number of atomic \textit{operations} performed by a kernel, denoted $O$, measures the total number of atomic operations across all SMs (e.g. one full-warp atomic instruction adds 32 to $O$).
$N_f^{(i)}$ and $N_c^{(i)}$ measures the number of FAO and CAS warp instructions executed on SM $i$, respectively.
The kernel time, denoted $T^{(i)}$, is measured as the number of clock cycles elapsed from a kernel launch to exit on SM $i$.

Quantities in Table~\ref{tab:derived-quantities} are derived from Table~\ref{tab:counters} using operational queuing theory, with our final goal of computing \emph{utilization}.
We approximate a few operational quantities used in the single-server model, because these quantities
cannot be measured directly with existing tools.
Such quantities include the average queue length $n^{(i)}$, which measures the parallelism of the atomic unit.
We approximate $n^{(i)}$ using $\hat{n}^{(i)}$, the average parallelism on SM $i$, by assuming that the average parallelism of the atomic unit is equal to that of the SM it belongs.
As NCU does not provide per-SM counters, we also assume that $e$ is the same for all SMs.
Once $\hat{n}^{(i)}$, $e$, and $c^{(i)}$ are obtained, $S^{(i)}$ can be calculated using the model parameters, and busy time is $B^{(i)} = N^{(i)}S^{(i)}$.
Obtaining utilization $U^{(i)}$ is then straightforward by using the measured kernel time $T^{(i)}$.
High values of utilization indicate that the shared memory atomic unit is a performance bottleneck.

\subsection{Profiling Tools: Implementation and Usage}\label{sec:tool}

We implemented the single-server model in two separate tools, one for obtaining model parameters, and one for profiling programs.
The first tool runs the microbenchmark %
to obtain a table of $S(n,e,c)$ for all valid integral values of $n$, $e$, and $c$.
This tool is only run once per model of GPU.

The second tool uses NVIDIA NSight Compute CLI and NVProf to profile the program and collect counter metrics listed in Table~\ref{tab:counters}.
After the program exits, the tool uses these metrics along with the model parameters collected by the first tool to instantiate the single-server model and compute shared-memory atomic utilization for each SM.

\section{Case Study: Image Histogram}\label{sec:histogram}

We now analyze two histogram kernels on the NVIDIA Titan V and A6000 GPUs.
We show how to observe data dependent performance, how the presence or absence of bottlenecks can be confirmed, and how to reason about performance optimizations and their effect.

An image histogram is
a basic operation %
used
in image processing to compute the frequency of each color in an image.
Histograms are an attractive use case of shared-memory atomics and have been used in prior work on modeling GPU performance~\cite{konstantinidisPracticalPerformanceModel2015,egielskiMassiveAtomicsMassive2014,gomez-lunaPerformanceModelingAtomic2013} because the input image can be split spatially into sub-problems to accumulate intermediate results in local histograms in shared memory, which can then be combined into a global histogram.

Listing~\ref{lst:hist} shows an abridged version of the first kernel we use to compute sub-histograms.
Its input %
is an image, supplied as a 1-D array of pixels.
In our experiments, images contain four 8-bit channels for red, green, blue and transparency (alpha) per pixel.
The input image and final output histogram (one per channel) are both stored in global memory, but each thread block computes local histograms in shared memory for its part of the image.

\begin{figure}
    \centering
    \includegraphics[width=0.9\linewidth]{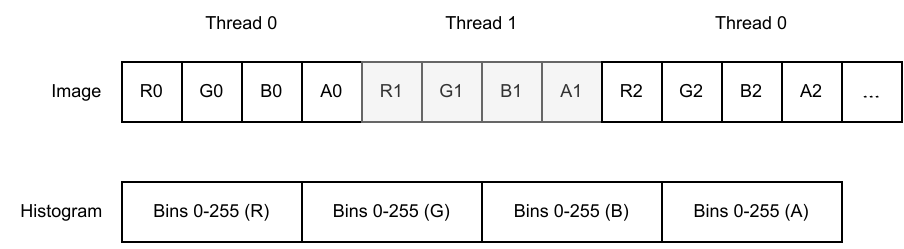}
    \caption{Histogram layout and assignment with 2 threads and 4 channels}
    \label{fig:hist-layout}
\end{figure}

\begin{lstlisting}[float,caption={Histogram kernel},label={lst:hist},language=C++]
    __global__ void hist(/*omitted*/) {
      __shared__ unsigned int smem[1024];
      // initialization omitted
      __syncthreads(); // start barrier
      for (int i = t; i < n; i += nt) {
        size_t in_base = i * channels;
        uint32_t offsets[4];
        // Collect channel value
        for (int c=0; c<channels; ++c) { 
          offsets[c] = in[in_base + c];
        }
        for (int c=0; c<channels; ++c) {
          // Write atomic
          int base = N_BINS * c;
          int* addr = &smem[base + offsets[c]];
          atomicAdd(addr, 1);
        }
      }
      __syncthreads(); // end barrier
      // output omitted
    }
    \end{lstlisting}

\begin{lstlisting}[float,caption={Histogram kernel reordered},label={lst:hist2},language=C++,firstnumber=13,escapeinside={(*}{*)}]
    // ...
    int chan = (c + tid % chans) % chans;(* \label{lst:hist2:tid} *)
    int base = N_BINS * chan;
    int* addr = &smem[base + offsets[chan]];
    atomicAdd(addr, 1);(* \label{lst:hist2:add} *)
    // ...
  \end{lstlisting}

Figure~\ref{fig:hist-layout} shows the memory layout of the image and the histogram along with the thread assignments.
Consecutive pixels are processed by consecutive threads.
Within each thread, the kernel processes assigned pixels sequentially, reading all channels of this pixel and updating the shared memory histogram.
After a thread block has processed all pixels assigned to it, the shared memory histogram is merged into the global memory histogram.%

When a large portion of the input image is a single color, multiple threads attempt to write into the same bins, potentially creating a bottleneck due to shared memory atomics.
To avoid this, as shown in Listing~\ref{lst:hist2}, we reorder the access to channels based on thread ID (Line~\ref{lst:hist2:tid}).
Now, when consecutive threads increment the histogram for the same color, their accesses  to the bins interleave and avoid bank conflicts, forming our second histogram kernel.

\subsection{Profiling the Kernels}\label{sec:profiling}

We computed the utilization of shared-memory atomic operations on two machines for both histogram kernels.
One machine has 2 Intel Xeon Silver 4208 CPU, 45GB RAM, and an NVIDIA Titan V GPU with 12GB RAM that contains 80 SMs.
The other has an AMD EPYC 7443P CPU, 256GB RAM, and an NVIDIA A6000 GPU with 48GB RAM that contains 84 SMs.
Both machines run CUDA 12.3.

We used two kinds of synthetic images with sizes of 32 pixels to 4 megapixels.
The first kind is monochromatic, with all pixels having the same value (solid).
The second kind contains uniformly random values for the channels of a pixel (uniform).
Maximum atomic contention occurs in solid images, while contention in the randomly colored images is low as atomics are randomly distributed.
Real-life images fall between these two extremes.

\begin{figure}
    \includegraphics[width=\columnwidth]{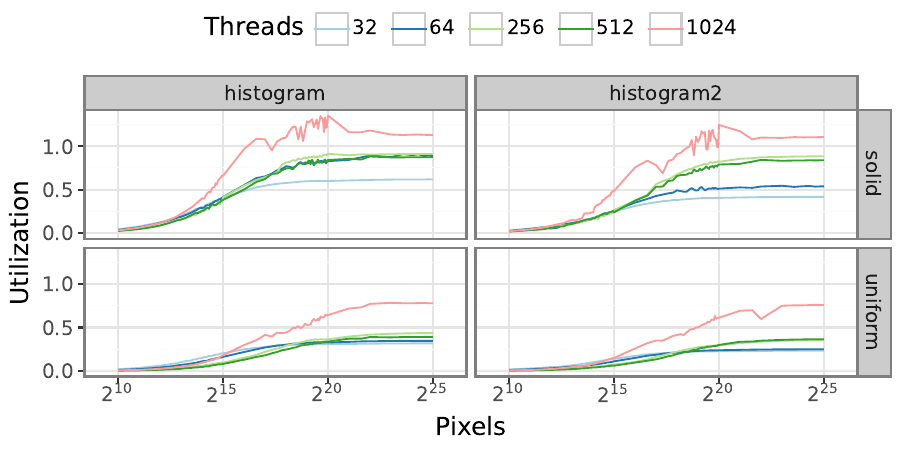}
    \centering
    \caption{Estimated shared memory atomic unit utilization of the original (\emph{histogram}) and updated (\emph{histogram2}) kernel on Titan V}\label{fig:hist}
\end{figure}

Figure~\ref{fig:hist} shows the estimated utilization of the histogram kernels on Titan V in our experiments, based on performance counters.
For small images, the low total number of atomics means the atomic unit has low utilization.
As the image size grows, the number of atomics increases increasing the utilization.
In solid color images, $e$ is 32, because all the atomics increment the same location.
On large inputs, the atomic units are fully utilized.
Our model reports an estimated utilization greater than 100\% on some measurements, because we  currently estimate $n$ and we suspect it is an overestimate leading to an overestimated utilization.
No GPU performance counter directly measures $n$ and we recommend GPU manufacturers add one.

For random color images, however, $e$ is 3 on average, and thus the atomic unit is never saturated with utilization peaking around 76\%.
Note, for a given image size, both monochromatic and random inputs execute the same number of atomic instructions and have the same parallelism ($n$), but differ only in the values of $e$.
The value of $n$ depends on thread block size, so for the same pixel size, fewer threads usually underutilize the shared memory unit.
This is not true for the 512 threads per block variant, and since the overall kernel time is the primary cause for this difference, we suspect there is overhead scheduling the 512 threads that gets exposed.

We also notice around the input size of $2^{20}$ pixels, the atomic utilization drops when 1024 threads are launched.
Upon inspecting performance counters, we find that the number of cache misses caused by reading the input increases significantly.
At this point, the bottleneck has likely shifted from atomic operations to global memory access whose latency cannot be hidden by the low number of threads.
This shift is unambiguously represented in our model's results and makes identifying performance bottlenecks for data-dependent kernels straightforward.

\begin{figure}
    \includegraphics[width=\columnwidth]{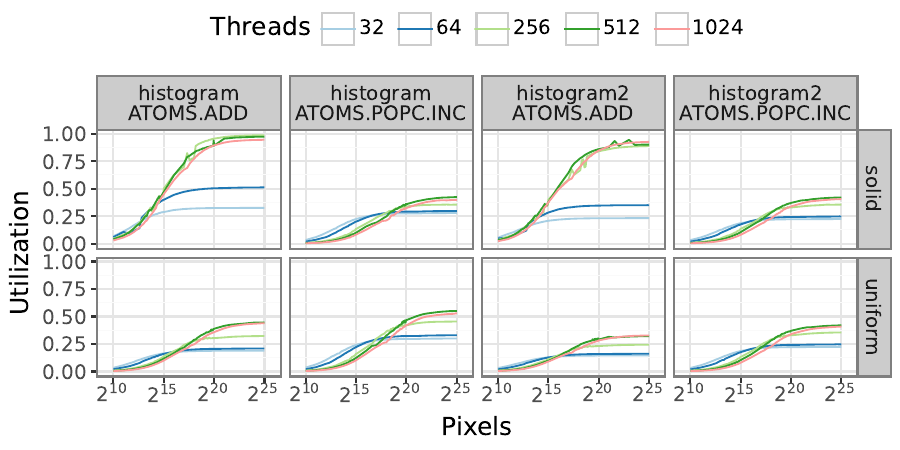}
    \centering
    \caption{Estimated shared memory atomic unit utilization for kernels on A6000}\label{fig:hist-algol}
\end{figure}

When compiling for the A6000, the CUDA compiler replaces the \texttt{ATOMS.ADD} instruction with the newer and cheaper \texttt{ATOMS.POPC.INC} since the result is not read.
We forced the compiler to generate \texttt{ATOMS.ADD} instead by performing a dummy read on the result (changing Listing~\ref{lst:hist2}, Line~\ref{lst:hist2:add}).
This results in two variants of \emph{histogram} and \emph{histogram2} with \texttt{ATOMS.ADD} instructions instead.

Figure~\ref{fig:hist-algol} shows the estimated atomic utilization on the A6000 GPU.
The \texttt{ATOMS.POPC.INC} instruction reduces utilization to 50\%, while the use of \texttt{ATOMS.ADD} keeps utilization near 100\%.
Modifying access order reduces atomic utilization for both instructions, but is more apparent in the \texttt{ATOMS.ADD} kernels.

\begin{figure}
    \includegraphics[width=\columnwidth]{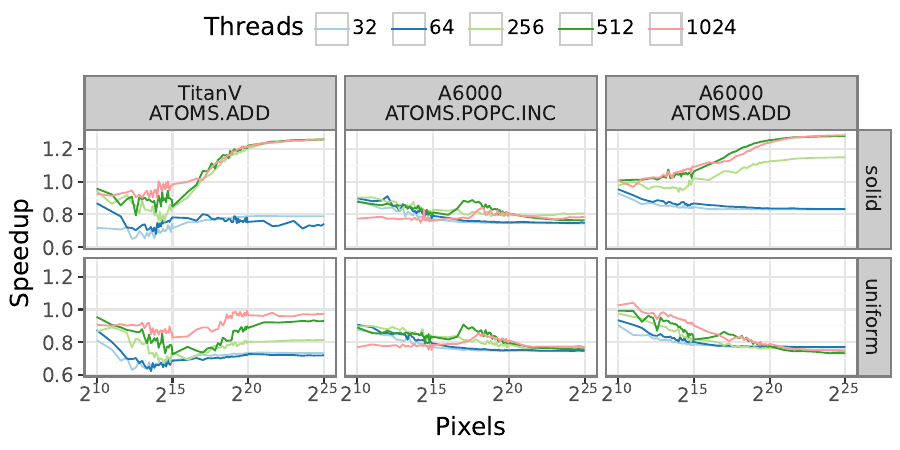}
    \caption{Histogram speedup on generated images}\label{fig:hist-speedup-by-pixels}
\end{figure}

Figure~\ref{fig:hist-speedup-by-pixels} shows the speedup for the channel reordered kernel over its corresponding unordered variant.
For both \texttt{ATOMS.ADD} kernels, reordering channels provides near 30\% speedup for large monochrome images when at least 512 threads are launched.
Random images suffer slowdown as the reorder operation adds some overhead.
Since Ampere's \texttt{ATOMS.POPC.INC} kernel's bottleneck was not shared memory atomics, the reordering slows down both kinds of images.

\section{Related Work}\label{sec:related}

Early work by \citet{zhang2011quantitative} microbenchmarked a variety of GPU architectural features, including the instruction pipeline, shared memory and global memory throughput on NVIDIA Tesla architecture, but not atomic operations.

\citet{gomez-lunaPerformanceModelingAtomic2013} and \citet{angGPUPERFORMANCEMODELING2016} modeled atomic operations on shared memory for the NVIDIA Fermi architecture which has a different implementation making their models unusable on current GPUs.

\citet{baghsorkhiAdaptivePerformanceModeling2010a} created an early analytical performance model of GPU programs using static program features and microbenchmarking results of the GPU.
It is intended to predict program performance statically, while our model measures utilization from a program run.

GPUMech~\citep{huangGPUMechGPUPerformance2014} used interval analysis to model the GPU but does not support atomic operations and requires detailed instruction traces and a functional simulator.

\citet{boyerImprovingGPUPerformance2013} focuses on the memory transfer between the host and the device memory.
\citet{gogolinska2018gpu} uses Petri nets to model arithmetic instructions, shared memory instructions, and global memory instructions.
\citet{lymDeLTAGPUPerformance2019a} models traffic between shared memory and hierarchy of caches in global memory in convolutional neural networks.
Neither supports atomic operations.

The Roofline model has been applied to GPUs in ~\citet{ding2019instruction} and \citet{lopes2017exploring} and is also the only analytical model available from the CUDA profiling tools~\cite{NsightKernelProfilingGuide} but does not provide analysis of atomic operations.

NVIDIA's CUDA Programming Manual provides throughput values for arithmetic instructions, some control instructions, and some memory instructions~\citep[Section~5]{cuda_c_programming_guide} but omits atomic instructions.
\citet{jiaDissectingNVIDIAVolta2018}
explores latency of shared memory atomic operations but does not examine detailed load-dependent behavior.

\citet{volkovUnderstandingLatencyHiding2016} compares multiple modeling methods on multiple GPUs (adding even more in~\cite{volkovMicrobenchmarkStudyGPU2018}), and provide a number of workloads that stress different part of the GPU and proposes a simple model for GPU performance based on Little's law, which is also used in operational analysis~\cite{denning_operational_1978}.
However, this work only considers extreme latency-bound and throughput-bound applications, not those in between and
does not include atomic operations.

\citet{arafaPPTGPUScalableGPU2019} is a hybrid approach with both static analysis and simulating using a GPU performance model, but does not consider atomic operations.

\citet{paiOperationalPerformanceModel2017a} used operational models to model the loads and stores in the GPU's memory hierarchy to predict the performance of graph programs on GPUs.
However, it only modeled ordinary loads and stores in global memory, and not atomic operations.

\citet{leeForecastingGPUPerformance2025} predicts GPU performance for deep learning applications using machine learning techniques.
It predicts utilization of compute and memory for deep learning operators, but does not include atomic units.

\section{Conclusion}\label{sec:conclusion}

We showed that utilization can be used to reason about data-dependent performance behavior.  %
Our profiling tool analyzes existing programs with no performance penalty over vendor-provided profiling tools.
Our model provides insight on shared-memory atomic unit utilization not provided by existing models or tools, using features already provided by the hardware.
Using case studies, we have demonstrated the model's ability to analyze similar implementations of the same algorithm.
In addition, intermediate parameters such as $n$ and $e$ provide additional insights in to program behavior.
Our work shows how GPU performance counters should evolve to handle data-dependent workloads.

Our method is also applicable to other GPU functional units, allowing programmers to identify bottlenecks in programs more accurately than existing methods. Though, it is ultimately limited by the performance counters a GPU provides and their granularity.

\begin{acks}
    We thank the anonymous reviewers for their suggestions which have made this paper better.
    This material is based upon work supported by the \grantsponsor{2144384}{National Science Foundation}{https://www.nsf.gov} under Grant No.~\grantnum[https://www.nsf.gov/awardsearch/showAward?AWD_ID=2144384]{nsf-grant-number}{2144384}.
    Any opinions, findings, and conclusions or recommendations expressed in this material are those of the author(s) and do not necessarily reflect the views of the National Science Foundation.
\end{acks}

\bibliography{Ref}


\begin{thebibliography}{39}

%%% the Web address.

\ifx \showCODEN    \undefined \def \showCODEN     #1{\unskip}     \fi
\ifx \showDOI      \undefined \def \showDOI       #1{#1}\fi
\ifx \showISBNx    \undefined \def \showISBNx     #1{\unskip}     \fi
\ifx \showISBNxiii \undefined \def \showISBNxiii  #1{\unskip}     \fi
\ifx \showISSN     \undefined \def \showISSN      #1{\unskip}     \fi
\ifx \showLCCN     \undefined \def \showLCCN      #1{\unskip}     \fi
\ifx \shownote     \undefined \def \shownote      #1{#1}          \fi
\ifx \showarticletitle \undefined \def \showarticletitle #1{#1}   \fi
\ifx \showURL      \undefined \def \showURL       {\relax}        \fi
% invisible to TeX
\providecommand\bibfield[2]{#2}
\providecommand\bibinfo[2]{#2}
\providecommand\natexlab[1]{#1}
\providecommand\showeprint[2][]{arXiv:#2}

\bibitem[CUD(2014)]%
        {CUDAProTipFilter2014}
 \bibinfo{year}{2014}\natexlab{}.
\newblock \bibinfo{title}{{{CUDA Pro Tip}}: {{Optimized Filtering}} with {{Warp-Aggregated Atomics}}}.
\newblock \bibinfo{howpublished}{https://developer.nvidia.com/blog/cuda-pro-tip-optimized-filtering-warp-aggregated-atomics/}.
\newblock


\bibitem[GPU(2015)]%
        {GPUProTipHistogram2015}
 \bibinfo{year}{2015}\natexlab{}.
\newblock \bibinfo{title}{{{GPU Pro Tip}}: {{Fast Histograms Using Shared Atomics}} on {{Maxwell}}}.
\newblock \bibinfo{howpublished}{https://developer.nvidia.com/blog/gpu-pro-tip-fast-histograms-using-shared-atomics-maxwell/}.
\newblock


\bibitem[Arafa et~al\mbox{.}(2019)]%
        {arafaPPTGPUScalableGPU2019}
\bibfield{author}{\bibinfo{person}{Yehia Arafa}, \bibinfo{person}{Abdel-Hameed~A. Badawy}, \bibinfo{person}{Gopinath Chennupati}, \bibinfo{person}{Nandakishore Santhi}, {and} \bibinfo{person}{Stephan Eidenbenz}.} \bibinfo{year}{2019}\natexlab{}.
\newblock \showarticletitle{{{PPT-GPU}}: {{Scalable GPU Performance Modeling}}}.
\newblock \bibinfo{journal}{\emph{IEEE Computer Architecture Letters}} \bibinfo{volume}{18}, \bibinfo{number}{1} (\bibinfo{date}{Jan.} \bibinfo{year}{2019}), \bibinfo{pages}{55--58}.
\newblock
\showISSN{1556-6064}
\urldef\tempurl%
\url{https://doi.org/10.1109/LCA.2019.2904497}
\showDOI{\tempurl}


\bibitem[Awan et~al\mbox{.}(2017)]%
        {awan2017depth}
\bibfield{author}{\bibinfo{person}{Ammar~Ahmad Awan}, \bibinfo{person}{Hari Subramoni}, {and} \bibinfo{person}{Dhabaleswar~K Panda}.} \bibinfo{year}{2017}\natexlab{}.
\newblock \showarticletitle{An in-depth performance characterization of CPU-and GPU-based DNN training on modern architectures}.
\newblock In \bibinfo{booktitle}{\emph{Proceedings of the Machine Learning on HPC Environments}}. \bibinfo{pages}{1--8}.
\newblock


\bibitem[Baghsorkhi et~al\mbox{.}(2010)]%
        {baghsorkhiAdaptivePerformanceModeling2010a}
\bibfield{author}{\bibinfo{person}{Sara~S. Baghsorkhi}, \bibinfo{person}{Matthieu Delahaye}, \bibinfo{person}{Sanjay~J. Patel}, \bibinfo{person}{William~D. Gropp}, {and} \bibinfo{person}{Wen-mei~W. Hwu}.} \bibinfo{year}{2010}\natexlab{}.
\newblock \showarticletitle{An Adaptive Performance Modeling Tool for {{GPU}} Architectures}. In \bibinfo{booktitle}{\emph{Proceedings of the 15th {{ACM SIGPLAN Symposium}} on {{Principles}} and {{Practice}} of {{Parallel Programming}}}} \emph{(\bibinfo{series}{{{PPoPP}} '10})}. \bibinfo{publisher}{{Association for Computing Machinery}}, \bibinfo{address}{{New York, NY, USA}}, \bibinfo{pages}{105--114}.
\newblock
\showISBNx{978-1-60558-877-3}
\urldef\tempurl%
\url{https://doi.org/10.1145/1693453.1693470}
\showDOI{\tempurl}


\bibitem[Boyer et~al\mbox{.}(2013)]%
        {boyerImprovingGPUPerformance2013}
\bibfield{author}{\bibinfo{person}{Michael Boyer}, \bibinfo{person}{Jiayuan Meng}, {and} \bibinfo{person}{Kalyan Kumaran}.} \bibinfo{year}{2013}\natexlab{}.
\newblock \showarticletitle{Improving {{GPU Performance Prediction}} with {{Data Transfer Modeling}}}. In \bibinfo{booktitle}{\emph{2013 {{IEEE International Symposium}} on {{Parallel}} \& {{Distributed Processing}}, {{Workshops}} and {{Phd Forum}}}}. \bibinfo{pages}{1097--1106}.
\newblock
\urldef\tempurl%
\url{https://doi.org/10.1109/IPDPSW.2013.236}
\showDOI{\tempurl}


\bibitem[{Corporation}(2022a)]%
        {cuda_c_programming_guide}
\bibfield{author}{\bibinfo{person}{NVIDIA {Corporation}}.} \bibinfo{year}{2022}\natexlab{a}.
\newblock \bibinfo{booktitle}{\emph{{CUDA} {C++} {Programming} {Guide}}}.
\newblock
\urldef\tempurl%
\url{https://docs.nvidia.com/cuda/archive/11.7.1/cuda-c-programming-guide/index.html}
\showURL{%
\tempurl}


\bibitem[{Corporation}(2022b)]%
        {cuda_cupti}
\bibfield{author}{\bibinfo{person}{NVIDIA {Corporation}}.} \bibinfo{year}{2022}\natexlab{b}.
\newblock \bibinfo{booktitle}{\emph{{{CUPTI}}}}.
\newblock
\urldef\tempurl%
\url{https://docs.nvidia.com/cupti/index.html}
\showURL{%
\tempurl}


\bibitem[{Corporation}(2022c)]%
        {NsightKernelProfilingGuide}
\bibfield{author}{\bibinfo{person}{{NVIDIA} {Corporation}}.} \bibinfo{year}{2022}\natexlab{c}.
\newblock \bibinfo{booktitle}{\emph{Kernel {{Profiling Guide}}}}.
\newblock {NVIDIA} {Corporation}.
\newblock
\urldef\tempurl%
\url{https://docs.nvidia.com/nsight-compute/2022.2/ProfilingGuide/index.html}
\showURL{%
\tempurl}


\bibitem[{Corporation}(2022d)]%
        {cuda_profiler_guide}
\bibfield{author}{\bibinfo{person}{{NVIDIA} {Corporation}}.} \bibinfo{year}{2022}\natexlab{d}.
\newblock \bibinfo{booktitle}{\emph{Profiler's {User} {Guide}}}.
\newblock {NVIDIA} {Corporation}.
\newblock
\urldef\tempurl%
\url{https://docs.nvidia.com/cuda/archive/11.7.1/profiler-users-guide/}
\showURL{%
\tempurl}


\bibitem[Denning and Buzen(1978)]%
        {denning_operational_1978}
\bibfield{author}{\bibinfo{person}{Peter~J. Denning} {and} \bibinfo{person}{Jeffrey~P. Buzen}.} \bibinfo{year}{1978}\natexlab{}.
\newblock \showarticletitle{The {Operational} {Analysis} of {Queueing} {Network} {Models}}.
\newblock \bibinfo{journal}{\emph{Comput. Surveys}} \bibinfo{volume}{10}, \bibinfo{number}{3} (\bibinfo{date}{Sept.} \bibinfo{year}{1978}), \bibinfo{pages}{225--261}.
\newblock
\showISSN{0360-0300}
\urldef\tempurl%
\url{https://doi.org/10.1145/356733.356735}
\showDOI{\tempurl}


\bibitem[Ding and Williams(2019)]%
        {ding2019instruction}
\bibfield{author}{\bibinfo{person}{Nan Ding} {and} \bibinfo{person}{Samuel Williams}.} \bibinfo{year}{2019}\natexlab{}.
\newblock \showarticletitle{An {{Instruction Roofline Model}} for {{GPUs}}}. In \bibinfo{booktitle}{\emph{2019 {{IEEE}}/{{ACM Performance Modeling}}, {{Benchmarking}} and {{Simulation}} of {{High Performance Computer Systems}} ({{PMBS}})}}. \bibinfo{pages}{7--18}.
\newblock
\urldef\tempurl%
\url{https://doi.org/10.1109/PMBS49563.2019.00007}
\showDOI{\tempurl}


\bibitem[Egielski et~al\mbox{.}(2014)]%
        {egielskiMassiveAtomicsMassive2014}
\bibfield{author}{\bibinfo{person}{Ian Egielski}, \bibinfo{person}{Jesse Huang}, {and} \bibinfo{person}{Eddy~Z Zhang}.} \bibinfo{year}{2014}\natexlab{}.
\newblock \showarticletitle{Massive {{Atomics}} for {{Massive Parallelism}} on {{GPUs}}}.
\newblock \bibinfo{journal}{\emph{ACM SIGPLAN}} \bibinfo{volume}{49}, \bibinfo{number}{11} (\bibinfo{date}{Nov.} \bibinfo{year}{2014}), \bibinfo{pages}{93--103}.
\newblock


\bibitem[Gogoli{\'n}ska et~al\mbox{.}(2018)]%
        {gogolinska2018gpu}
\bibfield{author}{\bibinfo{person}{Anna Gogoli{\'n}ska}, \bibinfo{person}{{\L}ukasz Mikulski}, {and} \bibinfo{person}{Marcin Pi{\k{a}}tkowski}.} \bibinfo{year}{2018}\natexlab{}.
\newblock \showarticletitle{GPU computations and memory access model based on Petri nets}.
\newblock In \bibinfo{booktitle}{\emph{Transactions on Petri Nets and Other Models of Concurrency XIII}}. \bibinfo{publisher}{Springer}, \bibinfo{pages}{136--157}.
\newblock


\bibitem[{G{\'o}mez-Luna} et~al\mbox{.}(2013)]%
        {gomez-lunaPerformanceModelingAtomic2013}
\bibfield{author}{\bibinfo{person}{Juan {G{\'o}mez-Luna}}, \bibinfo{person}{Jos{\'e}~Mar{\'i}a {Gonz{\'a}lez-Linares}}, \bibinfo{person}{Jos{\'e}~Ignacio Benavides~Ben{\'i}tez}, {and} \bibinfo{person}{Nicol{\'a}s Guil~Mata}.} \bibinfo{year}{2013}\natexlab{}.
\newblock \showarticletitle{Performance {{Modeling}} of {{Atomic Additions}} on {{GPU Scratchpad Memory}}}.
\newblock \bibinfo{journal}{\emph{IEEE Transactions on Parallel and Distributed Systems}} \bibinfo{volume}{24}, \bibinfo{number}{11} (\bibinfo{date}{Nov.} \bibinfo{year}{2013}), \bibinfo{pages}{2273--2282}.
\newblock
\showISSN{1558-2183}
\urldef\tempurl%
\url{https://doi.org/10.1109/TPDS.2012.319}
\showDOI{\tempurl}


\bibitem[Harris(2007)]%
        {harrisOptimizingParallelReduction2007}
\bibfield{author}{\bibinfo{person}{Mark Harris}.} \bibinfo{year}{2007}\natexlab{}.
\newblock \bibinfo{title}{Optimizing {{Parallel Reduction}} in {{CUDA}}}.
\newblock
\newblock
\urldef\tempurl%
\url{https://developer.download.nvidia.com/assets/cuda/files/reduction.pdf}
\showURL{%
\tempurl}


\bibitem[Hermann et~al\mbox{.}(2010)]%
        {hermann2010multi}
\bibfield{author}{\bibinfo{person}{Everton Hermann}, \bibinfo{person}{Bruno Raffin}, \bibinfo{person}{Fran{\c{c}}ois Faure}, \bibinfo{person}{Thierry Gautier}, {and} \bibinfo{person}{J{\'e}r{\'e}mie Allard}.} \bibinfo{year}{2010}\natexlab{}.
\newblock \showarticletitle{Multi-GPU and multi-CPU parallelization for interactive physics simulations}. In \bibinfo{booktitle}{\emph{European Conference on Parallel Processing}}. Springer, \bibinfo{pages}{235--246}.
\newblock


\bibitem[Huang et~al\mbox{.}(2014)]%
        {huangGPUMechGPUPerformance2014}
\bibfield{author}{\bibinfo{person}{Jen-Cheng Huang}, \bibinfo{person}{Joo~Hwan Lee}, \bibinfo{person}{Hyesoon Kim}, {and} \bibinfo{person}{Hsien-Hsin~S. Lee}.} \bibinfo{year}{2014}\natexlab{}.
\newblock \showarticletitle{{{GPUMech}}: {{GPU Performance Modeling Technique Based}} on {{Interval Analysis}}}. In \bibinfo{booktitle}{\emph{2014 47th {{Annual IEEE}}/{{ACM International Symposium}} on {{Microarchitecture}}}}. \bibinfo{pages}{268--279}.
\newblock
\showISSN{2379-3155}
\urldef\tempurl%
\url{https://doi.org/10.1109/MICRO.2014.59}
\showDOI{\tempurl}


\bibitem[Jespersen(2010)]%
        {jespersen2010acceleration}
\bibfield{author}{\bibinfo{person}{Dennis~C Jespersen}.} \bibinfo{year}{2010}\natexlab{}.
\newblock \showarticletitle{Acceleration of a CfD code with a GPU}.
\newblock \bibinfo{journal}{\emph{Scientific Programming}} \bibinfo{volume}{18}, \bibinfo{number}{3-4} (\bibinfo{year}{2010}), \bibinfo{pages}{193--201}.
\newblock


\bibitem[Jia et~al\mbox{.}(2018)]%
        {jiaDissectingNVIDIAVolta2018}
\bibfield{author}{\bibinfo{person}{Zhe Jia}, \bibinfo{person}{Marco Maggioni}, \bibinfo{person}{Benjamin Staiger}, {and} \bibinfo{person}{Daniele~P. Scarpazza}.} \bibinfo{year}{2018}\natexlab{}.
\newblock \bibinfo{title}{Dissecting the {{NVIDIA Volta GPU Architecture}} via {{Microbenchmarking}}}.
\newblock
\newblock
\urldef\tempurl%
\url{https://doi.org/10.48550/arXiv.1804.06826}
\showDOI{\tempurl}
\showeprint[arxiv]{1804.06826}~[cs]


\bibitem[Konstantinidis and Cotronis(2015)]%
        {konstantinidisPracticalPerformanceModel2015}
\bibfield{author}{\bibinfo{person}{Elias Konstantinidis} {and} \bibinfo{person}{Yiannis Cotronis}.} \bibinfo{year}{2015}\natexlab{}.
\newblock \showarticletitle{A {{Practical Performance Model}} for {{Compute}} and {{Memory Bound GPU Kernels}}}. In \bibinfo{booktitle}{\emph{2015 23rd {{Euromicro International Conference}} on {{Parallel}}, {{Distributed}}, and {{Network-Based Processing}}}}. \bibinfo{pages}{651--658}.
\newblock
\showISSN{2377-5750}
\urldef\tempurl%
\url{https://doi.org/10.1109/PDP.2015.51}
\showDOI{\tempurl}


\bibitem[Lee et~al\mbox{.}(2025)]%
        {leeForecastingGPUPerformance2025}
\bibfield{author}{\bibinfo{person}{Seonho Lee}, \bibinfo{person}{Amar Phanishayee}, {and} \bibinfo{person}{Divya Mahajan}.} \bibinfo{year}{2025}\natexlab{}.
\newblock \showarticletitle{Forecasting {{GPU Performance}} for {{Deep Learning Training}} and {{Inference}}}. In \bibinfo{booktitle}{\emph{Proceedings of the 30th {{ACM International Conference}} on {{Architectural Support}} for {{Programming Languages}} and {{Operating Systems}}, {{Volume}} 1}} \emph{(\bibinfo{series}{{{ASPLOS}} '25})}. \bibinfo{publisher}{Association for Computing Machinery}, \bibinfo{address}{New York, NY, USA}, \bibinfo{pages}{493--508}.
\newblock
\showISBNx{979-8-4007-0698-1}
\urldef\tempurl%
\url{https://doi.org/10.1145/3669940.3707265}
\showDOI{\tempurl}


\bibitem[Li(2016)]%
        {angGPUPERFORMANCEMODELING2016}
\bibfield{author}{\bibinfo{person}{Ang Li}.} \bibinfo{year}{2016}\natexlab{}.
\newblock \emph{\bibinfo{title}{{{GPU PERFORMANCE MODELING AND OPTIMIZATION}}}}.
\newblock Thesis. \bibinfo{school}{Eindhoven University of Technology}.
\newblock


\bibitem[Li et~al\mbox{.}(2017)]%
        {li2017survey}
\bibfield{author}{\bibinfo{person}{Zhen Li}, \bibinfo{person}{Yuqing Wang}, \bibinfo{person}{Tian Zhi}, {and} \bibinfo{person}{Tianshi Chen}.} \bibinfo{year}{2017}\natexlab{}.
\newblock \showarticletitle{A survey of neural network accelerators}.
\newblock \bibinfo{journal}{\emph{Frontiers of Computer Science}} \bibinfo{volume}{11}, \bibinfo{number}{5} (\bibinfo{year}{2017}), \bibinfo{pages}{746--761}.
\newblock


\bibitem[Liang et~al\mbox{.}(2018)]%
        {liang2018gpu}
\bibfield{author}{\bibinfo{person}{Jacky Liang}, \bibinfo{person}{Viktor Makoviychuk}, \bibinfo{person}{Ankur Handa}, \bibinfo{person}{Nuttapong Chentanez}, \bibinfo{person}{Miles Macklin}, {and} \bibinfo{person}{Dieter Fox}.} \bibinfo{year}{2018}\natexlab{}.
\newblock \showarticletitle{GPU-accelerated robotic simulation for distributed reinforcement learning}. In \bibinfo{booktitle}{\emph{Conference on Robot Learning}}. PMLR, \bibinfo{pages}{270--282}.
\newblock


\bibitem[Liu et~al\mbox{.}(2020)]%
        {liuWhyGPUsAre2020}
\bibfield{author}{\bibinfo{person}{Hongyuan Liu}, \bibinfo{person}{Sreepathi Pai}, {and} \bibinfo{person}{Adwait Jog}.} \bibinfo{year}{2020}\natexlab{}.
\newblock \showarticletitle{Why {{GPUs}} Are {{Slow}} at {{Executing NFAs}} and {{How}} to {{Make}} Them {{Faster}}}.
\newblock In \bibinfo{booktitle}{\emph{Proceedings of the {{Twenty-Fifth International Conference}} on {{Architectural Support}} for {{Programming Languages}} and {{Operating Systems}}}}. \bibinfo{publisher}{{Association for Computing Machinery}}, \bibinfo{address}{{New York, NY, USA}}, \bibinfo{pages}{251--265}.
\newblock
\showISBNx{978-1-4503-7102-5}


\bibitem[Lopes et~al\mbox{.}(2017)]%
        {lopes2017exploring}
\bibfield{author}{\bibinfo{person}{Andre Lopes}, \bibinfo{person}{frederico Pratas}, \bibinfo{person}{Leonel Sousa}, {and} \bibinfo{person}{Aleksandar Ilic}.} \bibinfo{year}{2017}\natexlab{}.
\newblock \showarticletitle{Exploring GPU performance, power and energy-efficiency bounds with Cache-aware Roofline Modeling}. In \bibinfo{booktitle}{\emph{2017 IEEE International Symposium on Performance Analysis of Systems and Software (ISPASS)}}. IEEE, \bibinfo{pages}{259--268}.
\newblock


\bibitem[Lym et~al\mbox{.}(2019)]%
        {lymDeLTAGPUPerformance2019a}
\bibfield{author}{\bibinfo{person}{Sangkug Lym}, \bibinfo{person}{Donghyuk Lee}, \bibinfo{person}{Mike O'Connor}, \bibinfo{person}{Niladrish Chatterjee}, {and} \bibinfo{person}{Mattan Erez}.} \bibinfo{year}{2019}\natexlab{}.
\newblock \showarticletitle{{{DeLTA}}: {{GPU Performance Model}} for {{Deep Learning Applications}} with {{In-Depth Memory System Traffic Analysis}}}. In \bibinfo{booktitle}{\emph{2019 {{IEEE International Symposium}} on {{Performance Analysis}} of {{Systems}} and {{Software}} ({{ISPASS}})}}. \bibinfo{pages}{293--303}.
\newblock
\urldef\tempurl%
\url{https://doi.org/10.1109/ISPASS.2019.00041}
\showDOI{\tempurl}


\bibitem[Navarro et~al\mbox{.}(2014)]%
        {navarro2014survey}
\bibfield{author}{\bibinfo{person}{Cristobal~A Navarro}, \bibinfo{person}{Nancy Hitschfeld-Kahler}, {and} \bibinfo{person}{Luis Mateu}.} \bibinfo{year}{2014}\natexlab{}.
\newblock \showarticletitle{A survey on parallel computing and its applications in data-parallel problems using GPU architectures}.
\newblock \bibinfo{journal}{\emph{Communications in Computational Physics}} \bibinfo{volume}{15}, \bibinfo{number}{2} (\bibinfo{year}{2014}), \bibinfo{pages}{285--329}.
\newblock


\bibitem[NVIDIA(2024)]%
        {cudacccl}
\bibfield{author}{\bibinfo{person}{NVIDIA}.} \bibinfo{year}{2024}\natexlab{}.
\newblock \bibinfo{title}{CUDA Core Compute Libraries (CCCL)}.
\newblock
\newblock
\urldef\tempurl%
\url{https://github.com/nvidia/cccl}
\showURL{%
\tempurl}


\bibitem[Nyland and Jones(2013)]%
        {nylandUnderstandingUsingAtomica}
\bibfield{author}{\bibinfo{person}{Lars Nyland} {and} \bibinfo{person}{Stephen Jones}.} \bibinfo{year}{2013}\natexlab{}.
\newblock \showarticletitle{Understanding and using atomic memory operations}. In \bibinfo{booktitle}{\emph{4th GPU Technology Conf.(GTC’13), March}}. \bibinfo{pages}{1--61}.
\newblock


\bibitem[Owens et~al\mbox{.}(2008)]%
        {owens2008gpu}
\bibfield{author}{\bibinfo{person}{John~D Owens}, \bibinfo{person}{Mike Houston}, \bibinfo{person}{David Luebke}, \bibinfo{person}{Simon Green}, \bibinfo{person}{John~E Stone}, {and} \bibinfo{person}{James~C Phillips}.} \bibinfo{year}{2008}\natexlab{}.
\newblock \showarticletitle{GPU computing}.
\newblock \bibinfo{journal}{\emph{Proc. IEEE}} \bibinfo{volume}{96}, \bibinfo{number}{5} (\bibinfo{year}{2008}), \bibinfo{pages}{879--899}.
\newblock


\bibitem[Pai et~al\mbox{.}(2017)]%
        {paiOperationalPerformanceModel2017a}
\bibfield{author}{\bibinfo{person}{Sreepathi Pai}, \bibinfo{person}{M~Amber Hassaan}, {and} \bibinfo{person}{Keshav Pingali}.} \bibinfo{year}{2017}\natexlab{}.
\newblock \showarticletitle{An {{Operational Performance Model}} of {{Breadth-First Search}}}. In \bibinfo{booktitle}{\emph{1st International Workshop on Architectures for Graph Processing (AGP)}}.
\newblock


\bibitem[Thomas(1976)]%
        {kleinrock1975theory}
\bibfield{author}{\bibinfo{person}{Marlin~U Thomas}.} \bibinfo{year}{1976}\natexlab{}.
\newblock \showarticletitle{Queueing systems. volume 1: Theory (leonard kleinrock)}.
\newblock \bibinfo{journal}{\emph{SIAM Rev.}} \bibinfo{volume}{18}, \bibinfo{number}{3} (\bibinfo{year}{1976}), \bibinfo{pages}{512--514}.
\newblock


\bibitem[Volkov(2016)]%
        {volkovUnderstandingLatencyHiding2016}
\bibfield{author}{\bibinfo{person}{Vasily Volkov}.} \bibinfo{year}{2016}\natexlab{}.
\newblock \emph{\bibinfo{title}{Understanding {{Latency Hiding}} on {{GPUs}}}}.
\newblock \bibinfo{thesistype}{Ph.\,D. Dissertation}. \bibinfo{school}{UC Berkeley}.
\newblock


\bibitem[Volkov(2018)]%
        {volkovMicrobenchmarkStudyGPU2018}
\bibfield{author}{\bibinfo{person}{Vasily Volkov}.} \bibinfo{year}{2018}\natexlab{}.
\newblock \showarticletitle{A Microbenchmark to Study {{GPU}} Performance Models}. In \bibinfo{booktitle}{\emph{Proceedings of the 23rd {{ACM SIGPLAN Symposium}} on {{Principles}} and {{Practice}} of {{Parallel Programming}}}} \emph{(\bibinfo{series}{{{PPoPP}} '18})}. \bibinfo{publisher}{{Association for Computing Machinery}}, \bibinfo{address}{{New York, NY, USA}}, \bibinfo{pages}{421--422}.
\newblock
\showISBNx{978-1-4503-4982-6}
\urldef\tempurl%
\url{https://doi.org/10.1145/3178487.3178536}
\showDOI{\tempurl}


\bibitem[Vouzis and Sahinidis(2011)]%
        {vouzis2011gpu}
\bibfield{author}{\bibinfo{person}{Panagiotis~D Vouzis} {and} \bibinfo{person}{Nikolaos~V Sahinidis}.} \bibinfo{year}{2011}\natexlab{}.
\newblock \showarticletitle{GPU-BLAST: using graphics processors to accelerate protein sequence alignment}.
\newblock \bibinfo{journal}{\emph{Bioinformatics}} \bibinfo{volume}{27}, \bibinfo{number}{2} (\bibinfo{year}{2011}), \bibinfo{pages}{182--188}.
\newblock


\bibitem[Wang et~al\mbox{.}(2010)]%
        {wang2010optimizing}
\bibfield{author}{\bibinfo{person}{Zhuowei Wang}, \bibinfo{person}{Xianbin Xu}, \bibinfo{person}{Wuqing Zhao}, \bibinfo{person}{Yuping Zhang}, {and} \bibinfo{person}{Shuibing He}.} \bibinfo{year}{2010}\natexlab{}.
\newblock \showarticletitle{Optimizing sparse matrix-vector multiplication on CUDA}. In \bibinfo{booktitle}{\emph{2010 2nd International Conference on Education Technology and Computer}}, Vol.~\bibinfo{volume}{4}. IEEE, \bibinfo{pages}{V4--109}.
\newblock


\bibitem[Zhang and Owens(2011)]%
        {zhang2011quantitative}
\bibfield{author}{\bibinfo{person}{Yao Zhang} {and} \bibinfo{person}{John~D Owens}.} \bibinfo{year}{2011}\natexlab{}.
\newblock \showarticletitle{A quantitative performance analysis model for GPU architectures}. In \bibinfo{booktitle}{\emph{2011 IEEE 17th international symposium on high performance computer architecture}}. IEEE, \bibinfo{pages}{382--393}.
\newblock


\end{thebibliography}

\end{document}